\documentclass[11pt]{article}

\usepackage[utf8]{inputenc}
\usepackage[T1]{fontenc}
\usepackage[margin=1in]{geometry}
\usepackage{amsmath,amsthm}
\usepackage{newtxtext}
\usepackage[varvw]{newtxmath}
\usepackage{booktabs}
\usepackage{microtype}
\usepackage{hyperref}
\usepackage[numbers,sort&compress]{natbib}
\usepackage{xcolor}
\usepackage{listings}
\lstdefinestyle{lean}{
  basicstyle=\ttfamily\small,
  columns=fullflexible,
  keepspaces=true,
  breaklines=true,
  literate=%
    {ℕ}{{$\mathbb{N}$}}1 {→}{{$\to$}}1 {∀}{{$\forall$}}1 {¬}{{$\neg$}}1
    {∈}{{$\in$}}1 {∧}{{$\wedge$}}1 {≤}{{$\le$}}1 {≠}{{$\ne$}}1
    {∨}{{$\vee$}}1 {⟨}{{$\langle$}}1 {⟩}{{$\rangle$}}1
    {∉}{{$\notin$}}1 {∃}{{$\exists$}}1
}

\hypersetup{
  colorlinks=true,
  linkcolor=blue!60!black,
  urlcolor=blue!50!black,
  citecolor=green!45!black,
}

\newtheorem{theorem}{Theorem}
\newtheorem{lemma}[theorem]{Lemma}
\newtheorem{corollary}[theorem]{Corollary}

\theoremstyle{definition}
\newtheorem{definition}[theorem]{Definition}

\theoremstyle{remark}

\newcommand{\leanname}[1]{%
  \texttt{\begingroup\def\_{\textunderscore\allowbreak}#1\endgroup}}
\newcommand{\mcso}{\mathrm{MCS}}
\newcommand{\Adj}{\mathsf{adj}}
\newcommand{\neigh}{\mathsf{neigh}}
\title{How rigid is maximum cardinality search? Stability and locality of
MCS orderings under edge updates, verified in Lean}

\author{Md Thoriqul Islam Thonmay\thanks{\href{mailto:mdthoriqulislam384@gmail.com}{\texttt{mdthoriqulislam384@gmail.com}}.
ORCID: \href{https://orcid.org/0009-0009-8317-6645}{0009-0009-8317-6645}}
\qquad Gregory Morse\thanks{\href{mailto:gregory.morse@live.com}{\texttt{gregory.morse@live.com}}.
ORCID: \href{https://orcid.org/0000-0002-0231-6557}{0000-0002-0231-6557}}\\[6pt]
\normalsize Department of Programming Languages and Compilers,\\
Faculty of Informatics, E\"otv\"os Lor\'and University, Budapest.}

\date{November 2026}

\begin{document}
\maketitle

\begin{abstract}
For which edge-modification sets $F$ do a graph $G$ and $G\triangle F$ share a
\emph{maximum cardinality search} (MCS) ordering? We prove that they do
whenever, along every proper common MCS prefix, each unchosen vertex's
neighbour count in the second graph exceeds its count in the first by zero or
one. In particular, inserting or deleting any \emph{matching}, or flipping a
single edge, preserves a common ordering, and under the same hypothesis every
common MCS prefix extends to a full common ordering without backtracking.
Neither part of the hypothesis can simply be dropped: inserting a triangle, a
$P_4$ or a $2P_3$ shows that the $+1$ bound cannot be relaxed to $+2$, and a
disjoint mixed insert/delete pair shows that the one-directional condition is
needed. For the canonical ordering (smallest-index
tie-breaking) under a single-edge update, an insertion can first change the
ordering only strictly between the endpoints' positions, and a deletion only at
the later endpoint; when no break fires, the ordering is unchanged. A
boolean-adjacency executable refines the abstract probe-then-repair update. All
of these results, including the four obstructions, are proved in Lean~4 over
Mathlib, re-checked by independent kernel checkers, and registered in the
Palomar registry. On paper we further show that the
first break of an insertion is located in $O((\deg(v)+1)\log n)$ time,
independent of the window length. We claim no worst-case asymptotic speedup over
recomputation.
\end{abstract}

\section{Introduction}
\label{sec:intro}

A \emph{maximum cardinality search} (MCS) numbers the vertices of a graph one at
a time, always choosing a vertex with the most already-chosen neighbours. Tarjan
and Yannakakis~\cite{tarjan1984mcs} introduced it as a \emph{simplification} of
the lexicographic breadth-first search (LexBFS) of Rose, Tarjan and
Lueker~\cite{rose1976algorithmic}: LexBFS compares the sets of chosen neighbours
lexicographically, MCS compares only their sizes. On \emph{chordal} graphs, the
reverse of an MCS ordering is a perfect elimination
ordering~\cite{tarjan1984mcs}, as Rose, Tarjan and Lueker showed for
LexBFS~\cite{rose1976algorithmic}; such orderings are the standard tool for
recognizing chordal graphs. Even on chordal graphs, the two ordering families are
incomparable~\cite{corneil2008unified}. Tarjan and Yannakakis number the vertices
from $n$ down to $1$ and prescribe no tie-break; we number forward and break ties
by smallest index throughout. Our question is dynamic: given an MCS ordering of a
graph, how much of it must change when one edge is inserted or deleted?

The answer is structural, with one asymptotic result, the probe below. We make
\textbf{no worst-case asymptotic separation claim}. In our unit-cost model an
update costs at most $n$ units for an insertion and $n+1$ for a deletion, the
same order as recomputing from scratch, and the static MCS of Tarjan and
Yannakakis already runs in linear time. Because the update provably returns the
from-scratch canonical ordering, our contribution is not a faster maintenance
algorithm but, to our knowledge, the first locality analysis of MCS orderings
under edge updates and the first machine-checked incremental repair of them
(Section~\ref{sec:related} records the survey behind this claim). What we prove
comes in four parts.
First, a \emph{stability theorem with obstructions} (Section~\ref{sec:stab})
that locates where two graphs stop sharing an MCS ordering. Second, the
\emph{locality} of a single-edge update (Section~\ref{sec:locality}): the open
window, or for a deletion the single position, where an update can first disturb
the ordering. Third, a \emph{sublinear first-break probe}
(Section~\ref{sec:probe}): the first break of an insertion is located in
$O((\deg(v)+1)\log n)$ time, independent of the window length, so a no-change
update can be certified in $o(n)$ time for sparse endpoints. Fourth, a
\emph{verified executable} (Section~\ref{sec:exec}) with a \emph{probe/pick
count} along its real control flow (Section~\ref{sec:cost}). Every theorem
except the probe bound is proved in Lean~4 and re-checked by an independent
external kernel checker. The probe bound (Theorem~\ref{thm:probe}) is proved on
paper: its plateau bound is machine-checked, and the data-structure argument is
on paper. Statements reported as
computer-search results (minimal obstruction sizes, patterns that never failed
up to a stated $n$) hold only for the sizes searched and are not theorems.
Section~\ref{sec:formalization} describes the artifact.

\subsection{Contributions}

\begin{enumerate}
\item \textbf{Stability theorem and obstructions.} Write $\mcso(\cdot)$ for the
\emph{set} of MCS orderings under arbitrary tie-breaking. For which
edge-modification sets $F$ is $\mcso(G)\cap\mcso(G\triangle F)\neq\varnothing$?
The intersection is nonempty whenever, along every proper common MCS prefix,
each unchosen vertex's count in $G\triangle F$ is at least its count in $G$ and
at most one more. That covers every
insertion-only or deletion-only matching and every single-edge flip. The
hypotheses cannot be relaxed: a triangle, a $P_4$, a $2P_3$ and a
\emph{disjoint} mixed pair each leave no common ordering. An online corollary
(\leanname{common\_order\_extends\_prefix}) ties the set-valued theory to the
constructive update: under the same hypothesis, any common MCS prefix extends to
a full common ordering without backtracking.
\item \textbf{Locality.} After a single-edge insertion, the ordering can first
change only inside the \emph{open} positional window between the two endpoints;
after a deletion, only at the later endpoint's position. When no break fires, the
ordering is \emph{literally unchanged}, not merely still valid.
\item \textbf{A sublinear first-break probe.} Per-value search trees over the
per-step data locate the first break of an insertion, or certify that there is
none, in $O((\deg(v)+1)\log n)$ time, independent of the window length, and
survive a no-break insertion at $O(\log n)$ cost. That beats the
naive $O(\text{window})\le O(n)$ scan roughly when
$(\deg(v)+1)\log n < \text{window}$, that is, for low-degree endpoints that lie
far apart in the ordering. A full update can still cost as much as
recomputation, and a break forces rebuilding the search trees; the gain is
certifying a no-change update in $o(n)$ for $\deg(v)=o(n/\log n)$.
\item \textbf{Verified executable and probe/pick count.} A boolean-adjacency
implementation refines the abstract update. An instrumented version counts
legality checks and pick steps along the real control flow: exactly $1$ for a
no-break deletion, and at most $|\mathrm{pos}(u)-\mathrm{pos}(v)|$ for a no-break
insertion.
\end{enumerate}

\paragraph{The cost model.} One cost unit is one legality check or one pick
step. A unit is itself superconstant work: in the naive list-matrix
implementation a pick filters the whole chosen list for each candidate, so one
unit can cost up to $\Theta(n^{3})$. The model therefore compares our update with
from-scratch recomputation \emph{within the same model}. It says nothing about
wall-clock parity with the linear-time static algorithm of Tarjan and
Yannakakis~\cite{tarjan1984mcs}. What we count is a \emph{probe/pick count}, not
a verified resource bound in the sense of time credits~\cite{gueneau2019formal},
and Section~\ref{sec:cost} says so again.

\section{Preliminaries}
\label{sec:prelim}

Fix $n\in\mathbb N$ and write $V_n=\mathrm{Fin}\,n$ for the vertex set. A graph
is a symmetric, loop-free binary relation $\Adj\colon V_n\times V_n\to\mathsf{Prop}$
equipped with proofs $\Adj\,u\,v\to\Adj\,v\,u$ and $\lnot\Adj\,v\,v$. For a set
$S\subseteq V_n$ and a vertex $w$, write $\neigh_G(S,w)$ for the number of
neighbours of $w$ inside $S$. A vertex $w\notin S$ is a \emph{legal MCS pick} for
the chosen set $S$ if $\neigh_G(S,w)$ is maximum over $V_n\setminus S$; it is a
\emph{canonical} pick if it is additionally smallest-index among the maxima. An
ordering $v_0,\dots,v_{n-1}$ is an MCS ordering if every prefix picks legally.

\begin{definition}[Canonical vs.\ set-valued]
\leanname{IsMCSOrdering} is the canonical predicate: at each step the picked
vertex is an MCS maximum with the smallest index. \leanname{IsMCSOrderingAny}
drops the smallest-index clause and asserts only that each step is
\emph{some} MCS maximum. The canonical ordering of a graph is unique
(\leanname{IsMCSOrdering\_unique}); $\mcso(G)$ denotes the \emph{set} of
orderings satisfying \leanname{IsMCSOrderingAny}.
\end{definition}

A single-edge update is modelled by $\leanname{FlipOf}\ G\ G'\ u\ v$: $G'$
agrees with $G$ except possibly on the pair $\{u,v\}$, with $u\neq v$, which
covers inserting or deleting that edge. The update procedure
$\leanname{insertUpdate}$ is a \emph{general flip procedure}: it keeps the
ordering up to the first break it finds between the endpoints (all of it if
there is none) and re-greedily rebuilds the suffix from there on the updated
graph, and it is correct for \emph{any} flip, not only an insertion. The deletion procedure is a
parallel probe-then-repair procedure that probes only the later endpoint's
position. The executable layer works over a concrete
boolean adjacency matrix $\leanname{Adj}\ n = \mathsf{List}\ (\mathsf{List}\
\mathsf{Bool})$ with a transport \leanname{toUGraph}.

A \emph{common MCS prefix} of $G$ and $G'$ is a duplicate-free partial ordering
whose every step is a legal MCS step in \emph{both} graphs. We write
$\leanname{IsCommonPrefix}\ G\ G'\ p$ for the predicate that $p$ is such a
prefix. The stability theorem of Section~\ref{sec:stab} requires its tracking
hypothesis only on the chosen sets of common MCS prefixes, and only for the
vertices not yet chosen, not on every subset.

\section{The stability theorem and obstructions}
\label{sec:stab}

The main structural question is: when do $G$ and $G'$ admit a \emph{common} MCS
ordering? The positive answer is stated under arbitrary tie-breaking, so it is
about the sets $\mcso(G)$ and $\mcso(G')$.

\begin{theorem}[Common ordering, \leanname{exists\_common\_ordering\_prefix}]
\label{thm:common}
Let $G,G'$ be graphs on $n$ vertices such that for every proper common MCS
prefix $p$ ($|p|<n$) of $G$ and $G'$ with chosen set $S$, and every vertex
$x\notin S$,
\[
\neigh_G(S,x)\ \le\ \neigh_{G'}(S,x)\ \le\ \neigh_G(S,x)+1.
\]
Then $\mcso(G)\cap\mcso(G')\neq\varnothing$.
\end{theorem}

The lower bound already forces the update to be one-directional. Every vertex is
a legal first pick, so every singleton $[y]$ is a common prefix, and the lower
bound at $S=\{y\}$ gives $\neigh_G(\{y\},x)\le\neigh_{G'}(\{y\},x)$ for all
$x\neq y$, that is, $E(G)\subseteq E(G')$ (for $n\le 1$ both graphs are
edgeless). Theorem~\ref{thm:common} is therefore
an insertion theorem; the deletion statements follow by exchanging the roles of
$G$ and $G'$, and the ``one-directional'' condition in the abstract is implied
by the hypothesis rather than added to it. The hypothesis is also not a
practical test: it ranges over all common prefixes, and enumerating those is the
same search that decides the conclusion. Theorem~\ref{thm:common} is a
structural sufficient condition; its usable content is the matching and flip
corollaries below and the online extension.

The proof is an induction that builds the common ordering one vertex at a time.
Its key step (\leanname{exists\_common\_pick}) is that every chosen set
$S\neq V_n$ satisfying the bounds admits a vertex that is an MCS maximum in
\emph{both} graphs: let $M$ be the maximum count in $G$ over $V_n\setminus S$; if
some $G$-maximizer gains $1$ in $G'$ it is maximal in both, and otherwise $M'=M$
and every $G$-maximizer works. The step compares unchosen vertices only, which
is why the bounds are needed only for $x\notin S$, and the induction visits only
common prefixes.

Both restrictions matter. Requiring the bounds for every subset $S$ and every
$x$ (\leanname{exists\_common\_ordering}, kept in the artifact) is equivalent to
$G'$ being $G$ plus a matching: $S=\{y\}$ forces $E(G)\subseteq E(G')$, and
$S=N_{G'}(x)$ forces $\deg_{G'}(x)\le\deg_G(x)+1$. Theorem~\ref{thm:common} is
strictly more general. Inserting a $P_3$ into the edgeless graph on three
vertices satisfies it: once one leaf is chosen, the centre is the only common
next pick, so no common prefix contains both leaves without the centre.
An exhaustive count over all insertion-only pairs $(G,F)$ with $F$ not a
matching shows the effect of each restriction: on $n=4$ vertices, $48$ pairs
satisfy the bounds for every $x$ along common prefixes and $100$ satisfy them
for unchosen $x$; on $n=5$ the counts are $1{,}020$ and $5{,}315$. In the same
count no pair satisfying the bounds lacks a common ordering, as the theorem
requires.

\begin{corollary}[Online form, \leanname{common\_order\_extends\_prefix}]
\label{cor:online}
The induction in Theorem~\ref{thm:common} is stronger than the theorem: under
the same hypothesis, with the bounds required only along proper common MCS
prefixes and only for unchosen vertices, any common MCS prefix extends to a full
common MCS ordering of both graphs. A
maintainer that, at each step, picks \emph{any} vertex that is maximum in both
graphs can therefore extend its prefix forever, with no lookahead and no
backtracking. This is a separate entry in the verified artifact, proved by strong
induction on the number of unplaced vertices. It also places the canonical
update of Section~\ref{sec:locality} in this picture: the prefix that update
keeps is a common MCS prefix, which under the hypothesis extends to a common
ordering; the update itself continues canonically in $G'$ only.
\end{corollary}

We apply Theorem~\ref{thm:common} to the update model. A \emph{matching} is a
graph in which every vertex has degree at most one.

\begin{corollary}[Matching insertions/deletions, \leanname{insert\_matching\_common\_order}, \leanname{delete\_matching\_common\_order}]
\label{cor:matching}
If $G'$ is obtained from $G$ by inserting a matching, or by deleting a matching,
then $\mcso(G)\cap\mcso(G')\neq\varnothing$.
\end{corollary}

\begin{corollary}[Single-edge flip, \leanname{flip\_common\_order}]
A single-edge flip always preserves a common MCS ordering: for any $u\neq v$
with $G'$ obtained from $G$ by flipping the edge $\{u,v\}$,
$\mcso(G)\cap\mcso(G')\neq\varnothing$.
\end{corollary}

The hypotheses of Theorem~\ref{thm:common} are precisely what one would try to
relax, and both fail. The ``$+1$'' bound cannot become ``$+2$'', and the
assumption that the update is one-directional (all inserts or all deletes) cannot
be dropped in general. Each negative result below is a concrete small instance,
certified in the artifact by a pruned search whose soundness is itself a theorem
and discharged by kernel decision on an explicit edge list
(Section~\ref{sec:formalization}).

\begin{theorem}[Certified obstructions: \leanname{triangle\_no\_common\_order}, \leanname{p4\_no\_common\_order}, \leanname{two\_p3\_no\_common\_order}, \leanname{mixed\_matching\_no\_common\_order}]
\label{thm:obstructions}
The following four edge-modification sets each destroy every common MCS ordering:

\begin{itemize}
\setlength\itemsep{2pt}
\item \textbf{Triangle} ($n=5$). Inserting a triangle into
$G=\{(0,2),(0,4),(1,2),(1,3)\}$ (adding $(2,3),(2,4),(3,4)$) leaves no common MCS
ordering: $\mcso(G)\cap\mcso(G')=\varnothing$. The triangle has maximum
$F$-degree~$2$.

\item \textbf{Path} ($n=6$). Inserting the triangle-free path
$2\text{--}5\text{--}4\text{--}3$ into the graph
$G=\{(0,3),\allowbreak(0,5),\allowbreak(1,2),\allowbreak(1,4),\allowbreak(2,3)\}$
leaves no common ordering. Unlike the triangle, the inserted $P_4$ is
triangle-free, so the obstruction does not reduce to the presence of a clique;
both cases have maximum $F$-degree~$2$.

\item \textbf{Two paths} ($n=8$). Inserting
$F=\{(0,1),\allowbreak(0,2),\allowbreak(3,4),\allowbreak(3,5)\}$ (a $2P_3$) into
$G=\{(1,5),\allowbreak(1,7),\allowbreak(2,4),\allowbreak(2,6),\allowbreak(4,5)\}$
leaves no common ordering.
With the triangle and $P_4$ it completes the three one-sided patterns used in
the Star--Matching discussion of Section~\ref{sec:conclusion}.

\item \textbf{Disjoint mixed pair} ($n=4$). Flipping the matching $\{01,23\}$ as
one insertion and one deletion (deleting $01$, inserting $23$) into
$G=\{(0,1),(0,3),(1,2)\}$ leaves no common ordering. The two flipped edges are
\emph{disjoint}, so the one-directional condition cannot be dropped for disjoint
mixed pairs.
\end{itemize}
\end{theorem}

Theorem~\ref{thm:obstructions} proves each obstruction on the stated witness
only. Exhaustive computer search over all hosts (the Python enumerator and the
C checker of Section~\ref{sec:formalization}) finds no smaller witness: the
smallest failing hosts have $n=5$ for the triangle, $n=6$ for $P_4$, $n=8$ for
$2P_3$ and $n=4$ for the disjoint mixed pair. In particular, inserting a $2P_3$
never destroys commonness on $n\le 7$ vertices, while $13{,}644$ of the $2^{24}$
hosts on $n=8$ fail. When the two flipped edges of a mixed pair share a vertex,
the same search found no failing host for $n\le 8$. These minimality and
no-failure facts are search results, not Lean theorems. No star-plus-matching
pattern
we tested ever failed: $P_3$, $K_{1,3}$ and $P_3+K_2$ over all hosts up to
$n=8$, and $K_{1,4}$, $K_{1,3}+K_2$ and $P_3+2K_2$ up to $n=7$.

The positive theorem is \emph{sufficient, not necessary}. On four vertices, take
$G=\{(1,2)\}$ and insert the two-edge path $F=\{(0,1),(0,2)\}$ (a $P_3$). The
common prefix $(3,1,2)$ leaves the centre $0$ unchosen with its count raised by
two, so the hypothesis fails, yet $G$ and $G+F$ still share an MCS ordering,
for instance $(0,1,2,3)$. A violation of the ``$+1$'' clause therefore does not
by itself destroy commonness. The exact boundary is posed as an open problem in
Section~\ref{sec:conclusion}.

\section{Locality of a single-edge update}
\label{sec:locality}

The canonical update is the \emph{probe-then-repair} procedure: scan the current
ordering for the first step whose old choice stops being the canonical MCS pick
in the updated graph, and re-greedily rebuild only the suffix from that point. Its
correctness is the local counterpart of the stability theorem.

\begin{theorem}[Update correctness, \leanname{init\_valid}, \leanname{insert\_update\_valid}, \leanname{delete\_update\_valid}, \leanname{greedySuffix\_full}, \leanname{update\_from\_prefix}]
Starting from a valid canonical ordering, a single insertion or deletion returns
a valid canonical MCS ordering of the updated graph
(\leanname{insert\_update\_valid}, \leanname{delete\_update\_valid}); the
empty-prefix regreedy initializes correctly (\leanname{init\_valid}),
regreedying a full prefix is the identity (\leanname{greedySuffix\_full}), and
regreedying a prefix of a valid canonical ordering of the old graph whose steps
are canonical in the updated graph yields a valid canonical ordering of the
updated graph (\leanname{update\_from\_prefix}).
\end{theorem}

\begin{theorem}[Update equals recompute, \leanname{insert\_update\_eq\_initOrder}, \leanname{delete\_update\_eq\_initOrder}, \leanname{IsMCSOrdering\_unique}]
The returned ordering is exactly the from-scratch canonical ordering of the
updated graph; hence, by uniqueness of the canonical ordering, the maintained
state is the canonical ordering, not just some valid one.
\end{theorem}

The locality of the probe is where the algorithm gains over recomputation.

\begin{theorem}[Insertion window, \leanname{legal\_upto\_earlierPos}, \leanname{legal\_after\_laterPos}, \leanname{insert\_break\_iff}]
\label{thm:insert-window}
For an insertion of $\{u,v\}$ with $u$ earlier than $v$ in the canonical order,
every position up to $u$'s position and every position after $v$'s position keeps
its old choice canonical in $G'$ with respect to the old prefix; inside the open window
$(\mathrm{pos}(u),\mathrm{pos}(v))$, a step breaks exactly when the later
endpoint $v$ now beats the old choice at that step.
\end{theorem}

\begin{proof}
The update adds the single edge $\{u,v\}$. For a prefix $S$ at or before
$\mathrm{pos}(u)$, $S$ contains neither endpoint, so every candidate count is
unchanged and the old pick stays canonical. For $S$ after $\mathrm{pos}(v)$, both
endpoints lie in $S$, so they are not candidates and the counts of the remaining
candidates from $S$ are identical in $G$ and $G'$; the old pick stays canonical.
For $S$ strictly inside the window, $u\in S$ and $v\notin S$, so $v$ is the only
candidate whose count changes, rising by $1$ (the new edge to $u$), and it
breaks the step exactly when it overtakes the old pick.
\end{proof}

\begin{theorem}[Deletion single-point, \leanname{legal\_upto\_earlierPos}, \leanname{delete\_legal\_between}, \leanname{legal\_after\_laterPos}]
\label{thm:delete-point}
For a deletion of $\{u,v\}$, only the later endpoint's own position can be the
first break; positions strictly between the endpoints keep their old choice
canonical in $G'$.
\end{theorem}

\begin{proof}
The update removes $\{u,v\}$. For a prefix $S$ at or before $\mathrm{pos}(u)$,
neither endpoint is in $S$, so no count from $S$ changes and the old pick stays
canonical. For $S$ after $\mathrm{pos}(v)$, both endpoints are in $S$ and are
not candidates, so the remaining counts are unchanged
(\leanname{legal\_after\_laterPos}). For $S$ strictly between the endpoints,
$u\in S$ but $v\notin S$: the count of $v$ drops by $1$ (it loses neighbour $u$),
while every other candidate's count is unchanged. Since $v$ is not yet the pick,
the old pick $w$ has $c_w\ge c_v$ before the drop and hence $c_w>c_v-1$ after it,
so $w$ remains the maximum and canonical. Thus the first break can only occur at
$\mathrm{pos}(v)$, where $v$ is itself the pick and its count drops.
\end{proof}

\begin{lemma}[No break at the later endpoint, \leanname{no\_break\_at\_laterPos}]
\label{lem:no-break-later}
An insertion never breaks at the later endpoint's own position; the window is
the open interval $(\mathrm{pos}(u),\mathrm{pos}(v))$, not the half-open
$(\mathrm{pos}(u),\mathrm{pos}(v)]$.
\end{lemma}

\begin{proof}
At step $\mathrm{pos}(v)$ the old pick is $v$ itself. Its count rises by $1$
(because $u$ is already chosen), and every other unchosen count is unchanged. So
$v$ becomes the \emph{strict} maximum and stays canonical.
\end{proof}

\begin{lemma}[Old-graph break criterion, \leanname{insert\_break\_iff\_old}]
\label{lem:old-break}
Inside the open window, with $w_j$ the old pick at step $j$ and $M_j$ the old
maximum count, the step breaks exactly when
\[
c_v(j)=M_j \quad\text{or}\quad \bigl(c_v(j)=M_j-1 \text{ and } v<w_j\bigr),
\]
where $c_v(j)=\neigh_G(\mathrm{prefix}_j, v)$ is the count in the \emph{old}
graph.
\end{lemma}

\begin{proof}
At a step $j$ in the open window, only $v$'s count changes, from $c_v(j)$ to
$c_v(j)+1$, and $c_v(j)\le M_j$. The step breaks exactly when $v$'s new count
exceeds $M_j$ (i.e.\ $c_v(j)=M_j$), or equals $M_j$ with a smaller index than the
old pick (i.e.\ $c_v(j)=M_j-1$ and $v<w_j$).
\end{proof}

Both lemmas are comparator theorems of the artifact. The first justifies
shrinking the probe window by one position; the executable does not exploit
this and still checks $\mathrm{pos}(v)$ (Section~\ref{sec:cost}). The second
makes the probe a question about old-graph data only, which
Section~\ref{sec:probe} exploits.

The update equals recompute, so the statement that the update is the identity
exactly when the ordering is already canonical is a tautology given the
update-equals-recompute theorem; the artifact retains it as an entry
(Section~\ref{sec:formalization}) but we do not headline it.

\section{A sublinear first-break probe}
\label{sec:probe}

The insertion probe asks: is there a step $j$ in the open window
$(\mathrm{pos}(u),\mathrm{pos}(v))$ at which the old pick stops being canonical
in the updated graph? By Lemma~\ref{lem:old-break} this is a question about the
old-graph data $(c_v(j), M_j, w_j)$. We prove the probe is answerable in
$O((\deg(v)+1)\log n)$ time, independent of the window length.

\begin{theorem}[Sublinear first-break probe]
\label{thm:probe}
Store the canonical order together with the per-step data $(M_j, w_j)$ and, for
each value $m$, a balanced search tree over $\mathrm{Pos}[m]=\{j : M_j=m\}$,
keyed by position and augmented with the maximum of $w_j$ in each subtree. For
an insertion of $\{u,v\}$, the \emph{first} position of the open window at which
the old ordering breaks, or the fact that there is none, is computed in
$O((\deg_G(v)+1)\log n)$ time, independent of the window length. After a
no-break insertion, $(w_j)$ is unchanged and $(M_j)$ changes only at
$j=\mathrm{pos}(v)$, by $+1$, so the structure is updated in $O(\log n)$ time.
Building it takes $O(n\log n)$ time and $O(n)$ space once the canonical
ordering and its per-step data are known; computing those takes
$O((n+m)\log n)$ time with a heap per count value for the smallest-index
tie-break, against $O(n+m)$ for the arbitrary tie-break of~\cite{tarjan1984mcs}.
\end{theorem}

\begin{proof}
Two facts are machine-checked in the artifact. First, the count
$c_v(j)=G.\neigh(\mathsf{take}\,j,\;v)$ is a step function: adding one vertex
changes it by $0$ or $1$, exactly per adjacency
(\leanname{neigh\_take\_succ}), and its exact net change over an interval
$[j,k)$ equals the number of neighbours of $v$ in that interval
(\leanname{neigh\_take\_sub\_eq\_count}). Consequently $c_v$ takes at most
$\deg_G(v)+1$ distinct values across the window
(\leanname{neigh\_take\_image\_card\_le}). This plateau bound is the
machine-checked part; the tree queries, the descent and the plateau scan below
are on paper. The plateau boundaries are the window positions just after a neighbour
of $v$, so sorting the positions of $v$'s neighbours yields them, and the initial
value $c_v$ at the window start, in $O((\deg_G(v)+1)\log n)$ time. On each
constant piece $[l,r)$ with $c_v(j)=c$, the criterion of
Lemma~\ref{lem:old-break} asks for the first $j\in[l,r)$ with $M_j=c$, or with
$M_j=c+1$ and $w_j>v$. The first is the successor of $l$ in the tree of
$\mathrm{Pos}[c]$; the second is found by descending the tree of
$\mathrm{Pos}[c+1]$ inside $[l,r)$, skipping every subtree whose maximum $w$ is
at most $v$. Both cost $O(\log n)$. The pieces are scanned in window order and
the scan stops at the first piece that yields a position; the earlier of its at
most two candidates is the first break. With at most $\deg_G(v)+1$ pieces the
total is $O((\deg_G(v)+1)\log n)$. After a no-break insertion only $M_{\mathrm{pos}(v)}$
changes, so $\mathrm{pos}(v)$ moves from the tree of $\mathrm{Pos}[M]$ to that of
$\mathrm{Pos}[M+1]$ in $O(\log n)$. The
win regime: the probe beats the naive $O(\text{window})\le O(n)$ scan roughly
when $(\deg(v)+1)\log n < \text{window}$, that is, for low-degree endpoints that lie
far apart in the ordering. A full update after a break can still cost as much
as recomputation, plus rebuilding the trees; the gain is certifying a no-change
update in $o(n)$ for $\deg_G(v)=o(n/\log n)$.
\end{proof}

The break predicate is not monotone in $j$. For example, on $n=4$ with
$G=\{1\text{--}2\}$ and canonical order $(0,1,2,3)$, insert $\{0,3\}$: the open
window is $\{1,2\}$ and the step function $j\mapsto[\text{break at }j]$ is
$(\text{yes},\text{no})$, so a break is followed by a non-break (no graph on
three vertices shows this). This rules out a naive
binary search on a single key, but the multi-plateau structure above is the
right shape for the range queries. The sublinear bound is proved here on paper
(the plateaus are machine-checked); it is not itself a Lean theorem, since the
artifact has no formal cost model. As a sanity check, the artifact script
\leanname{scripts/probe\_check.py} compares the probe with from-scratch
recomputation on $35{,}856$ random insertions ($n\le 13$): the reported first
break equals the first changed position in every case, and on all $13{,}077$
no-break insertions $(w_j)$ is unchanged and $(M_j)$ changes exactly at
$\mathrm{pos}(v)$, by $+1$.

\paragraph{Deletions.} By Theorem~\ref{thm:delete-point} a deletion can first
break only at $\mathrm{pos}(v)$, so \emph{locating} the break is free.
\emph{Deciding} whether it fires is not determined by the stored data
$(M_j,w_j)$: on three vertices, the hosts $\{01,02\}$ and $\{01,12\}$ have the
same canonical order $(0,1,2)$ and the same $(M_j)=(0,1,1)$, yet deleting $01$
changes the ordering of the first and not of the second. With per-step
runner-up data, $R_j$ the largest count among unchosen vertices other than
$w_j$ and $r_j$ the smallest vertex attaining it, the step at
$j=\mathrm{pos}(v)$ breaks iff $R_j=M_j$, or $R_j=M_j-1$ and $r_j<v$, an $O(1)$
test (confirmed against recomputation on $35{,}925$ random deletions). Keeping
$(R_j,r_j)$ current is the obstacle: a no-break deletion can change runner-up
data at other steps, and did so in $3{,}477$ of $7{,}156$ random no-break
deletions. A sublinear maintained deletion probe is therefore open
(Section~\ref{sec:conclusion}).

\section{Verified executable}
\label{sec:exec}

To rule out the gap where a ``specification'' is proved but nothing executable
implements it, the development carries a concrete boolean-adjacency
implementation and proves it refines the abstract update.

\begin{theorem}[Executable refinement, \leanname{exec\_insert\_update\_eq}, \leanname{exec\_delete\_update\_eq}, \leanname{setEdge\_preserves\_symm}, \leanname{setEdge\_preserves\_loopless}, \leanname{mcsOrder\_eq\_greedySuffix}]
\label{thm:exec-refine}
Let $a,a'$ be symmetric, loop-free boolean adjacency matrices with $|a'|=n$,
whose graphs $G=\leanname{toUGraph}\,a$ and $G'=\leanname{toUGraph}\,a'$ are
related by a flip of $\{u,v\}$ ($\leanname{FlipOf}\ G\ G'\ u\ v$), and let $o$ be
the canonical MCS ordering of the \emph{old} graph $G$. Then
\[
\text{\leanname{execInsertUpdate}}\ a'\ (\mathrm{map\ val}\ o)\ u\ v
  = \mathrm{map\ val}\ (\text{\leanname{insertUpdate}}\ G'\ o\ u\ v),
\]
and likewise $\leanname{execDeleteUpdate}$ equals $\leanname{deleteUpdate}$ when
$\{u,v\}$ is an edge of $a$ and not of $a'$. Setting or clearing the edge
$\{u,v\}$ of a well-shaped symmetric matrix preserves symmetry, and preserves
loop-freeness when $u\neq v$;
the executable from-scratch MCS of a matrix is the value list of a canonical MCS
ordering of its graph.
\end{theorem}

The artifact additionally carries theorems that certify \emph{recomputation},
not the dynamic update: they state that recomputing the MCS of the flipped
matrix from scratch is valid. The dynamic update is certified by the refinement
theorem above (Section~\ref{sec:formalization}).

\section{Probe/pick count along the real control flow}
\label{sec:cost}

An instrumented variant of the executable update counts, at each step, one unit
for a legality check and one unit for a pick step, and returns the count together
with the ordering. Its first projection is provably the plain update
(Section~\ref{sec:formalization}), so the counts below attach to the update that
is actually run.

\begin{theorem}[Probe/pick counts, \leanname{exec\_delete\_cost\_of\_no\_break}, \leanname{exec\_insert\_cost\_of\_no\_break}, \leanname{exec\_insert\_cost\_of\_break}, \leanname{exec\_delete\_cost\_of\_break}, \leanname{exec\_insert\_cost\_le}, \leanname{exec\_delete\_cost\_le}]
\label{thm:cost}
Let $n=|V|$. In the unit-cost model, where one unit is one legality check or
one pick step, the following hold.
\begin{enumerate}
\setlength\itemsep{2pt}
\item \textbf{No-break deletion}: if a deletion's probe finds no break, the
count is \emph{exactly} $1$ unit.

\item \textbf{No-break insertion}: if an insertion's probe finds no break, the
count is at most the window size $|\mathrm{pos}(u)-\mathrm{pos}(v)|$ for the two
endpoints.

\item \textbf{Break cost}: a break at position $k$ costs the probe plus at most
$n-k$ regreedy steps.

\item \textbf{Worst case}: at most $n$ units for insertion (given an ordering of
length $n$), and at most $n+1$ for deletion. Both are within one unit of
from-scratch recomputation in this model, which costs $n$ pick steps.
\end{enumerate}
\end{theorem}

The two most useful statements are the first two: a no-break deletion costs one
check, a no-break insertion costs a probe of the window. These are the
operational content of the locality results of Section~\ref{sec:locality}.
The insertion bound $|\mathrm{pos}(u)-\mathrm{pos}(v)|$ counts the open window
plus the later endpoint's own position, which the executable probe also checks;
by Lemma~\ref{lem:no-break-later} that check never fires, so the proved bound is
one unit above the open-window optimum.

A cost unit is superconstant in the naive list-matrix implementation: an
adjacency lookup indexes a list-of-lists at $O(n)$, a cardinality check costs
$O(|\text{chosen}|\cdot n)$, and a pick calls it for every remaining candidate,
so one pick costs $O(|\text{rem}|\cdot|\text{chosen}|\cdot n)$, up to
$\Theta(n^{3})$ when both lists have length $\Theta(n)$. The count is therefore
a \emph{probe/pick count}, weaker than a verified resource analysis in the sense of time
credits~\cite{gueneau2019formal}; it compares the update against from-scratch
recomputation within the same model and says nothing about wall-clock parity with
the linear-time static algorithm.

\section{The Lean formalization}
\label{sec:formalization}

The development is a Lean~4 / Mathlib project, organized as a Palomar registry
challenge and \emph{registered} as Palomar entry
\leanname{PALOMAR-2026-10-04-000009}.\footnote{Entry:
\url{https://palomar-registry.org/entry.html?id=PALOMAR-2026-10-04-000009&version=3};\newline
record: \url{https://data.palomar-registry.org/entries/PALOMAR-2026-10-04-000009-v3.json}.}
Version~3 is registered from commit \texttt{ab521f2}: status
\texttt{registered}, review outcome \texttt{neutral} with no review warnings,
trust level \texttt{high}. Its one trust reason is ``Challenge exceeds the
preferred audit surface'' (742 lines). The commit pinned below differs from
\texttt{ab521f2} only in the axiom audit (\texttt{AxiomAudit.lean} and its CI
script) and the README, so version~3 carries exactly the statements of this
paper, and its external kernel re-checks (\texttt{nanoda}, \texttt{con-ron})
cover them. The registry records
the source under the repository's former name, \texttt{thonmay/thesis-lean},
which redirects to the current one. Version~3 supersedes two earlier versions.
Version~2, from the ancestor commit \texttt{c4c2c0f}, requires the count bounds
of Theorem~\ref{thm:common} and Corollary~\ref{cor:online} for every vertex
rather than only for unchosen ones. Version~1, from the ancestor commit
\texttt{a872fdd}, has $37$ of the $41$ comparator theorems and states the online
corollary under every-subset bounds. Registration is a preservation and indexing service,
not a review endorsement.

A module \leanname{Challenge} poses the headline statements, with explicit
\texttt{sorry} bodies and canonical definitions in Mathlib-only code; a module
\leanname{Solution} proves them. Both are generated from a single source of
truth, so every posed statement is textually identical to its proved
counterpart.
The challenge/solution comparator lists $41$ theorem names. By the section in
which each is first named: $1$ in the introduction, $1$ in
Section~\ref{sec:prelim}, $13$ in Section~\ref{sec:locality}, $9$ in
Section~\ref{sec:stab}, $5$ in Section~\ref{sec:exec}, $6$ in
Section~\ref{sec:cost}, and $6$ only in the formalization section below. The
table groups them by the section where they are \emph{stated}; the two names
that first appear early (\leanname{common\_order\_extends\_prefix} and
\leanname{IsMCSOrdering\_unique}) are stated in Sections~\ref{sec:stab}
and~\ref{sec:locality}. They fall into these categories:

\begin{center}
\begin{tabular}{lrl}
\toprule
Category & count & Where stated \\
\midrule
update correctness & $5$ & \ref{sec:locality} \\
uniqueness / update = recompute & $3$ & \ref{sec:locality} \\
locality (incl.\ the two sharpening lemmas) & $6$ & \ref{sec:locality} \\
executable bridge & $3$ & \ref{sec:exec} \\
executable refinement & $2$ & \ref{sec:exec} \\
probe/pick count & $6$ & \ref{sec:cost} \\
stability (matching, flip, prefix extension) & $5$ & \ref{sec:stab} \\
every-subset special case & $1$ & \ref{sec:stab} \\
certified obstructions (triangle, $P_4$, $2P_3$, mixed) & $4$ & \ref{sec:stab} \\
recomputation certs., tautologies, projection links & $6$ & \ref{sec:formalization} \\
\midrule
total & $41$ & \\
\bottomrule
\end{tabular}
\end{center}

Beyond the $41$ comparator names, the development carries auxiliary and
proof-engineering entries that the mathematical sections reference but do not
headline: the key-step lemma \leanname{exists\_common\_pick}; the pruned-search
soundness lemmas \leanname{no\_common\_order\_of\_search} and
\leanname{commonSearch\_of\_common}; and the step-function facts
\leanname{neigh\_take\_succ}, \leanname{neigh\_take\_sub\_eq\_count} and
\leanname{neigh\_take\_image\_card\_le} that underpin the sublinear probe of
Section~\ref{sec:probe}.

The comparator's last row collects entries that certify the recomputation layer
rather than the dynamic update: the tautological identity entries
\leanname{insert\_update\_eq\_self\_iff} and
\leanname{delete\_update\_eq\_self\_iff}, the recomputation-certifying entries
\leanname{exec\_insert\_recompute\_valid} and
\leanname{exec\_delete\_recompute\_valid}, and the first-projection links
\leanname{exec\_insert\_update\_c\_fst} and
\leanname{exec\_delete\_update\_c\_fst} that tie the instrumented count to the
plain update.

Every theorem audits to only the three standard Lean axioms (\texttt{propext},
\texttt{Quot.sound}, \texttt{Classical.choice}): there is no \texttt{sorry}
outside \leanname{Challenge}, and no \leanname{native\_decide}. An
\texttt{AxiomAudit} module prints the axioms of each comparator theorem and of
the six auxiliary lemmas above, enforced as a CI step; a definitions-sync check enforces byte-identity of the canonical
definitions; and an external kernel checker (\texttt{leanchecker}) re-validates
each of the $20$ source modules. The certified obstructions are independently cross-checked by the
registry kernel re-checks (\texttt{nanoda} and \texttt{con-ron}) and by a separate
exhaustive Python enumerator that reproduces each witness edge list exactly.

\paragraph{Artifact and reproducibility.} The Lean development, its comparator,
the axiom audit and the empirical verification scripts are public at
\url{https://github.com/thonmay/mcs-lean} under the Apache-2.0 license, pinned
to the annotated tag \leanname{mcs-arxiv-v10} (commit \texttt{0d0ddff}), which is
also published as a GitHub Release and archived by Software
Heritage.\footnote{Software Heritage revision identifier:\newline
\texttt{swh:1:rev:0d0ddffbcef5817c50a073a350428edd2da5bfb5};\newline
origin: \url{https://github.com/thonmay/mcs-lean}.} It differs from
\texttt{ab521f2} only in the axiom audit (\texttt{AxiomAudit.lean} and
\texttt{scripts/check\_axioms.sh}), which also covers the six auxiliary lemmas
cited above, and in the README; the posed and proved statements are
byte-identical. Palomar version~3 is registered from \texttt{ab521f2} (see
above). The README on the \texttt{master} branch maps every theorem
named in this paper to its Lean source file and lists the exact build and
verification commands (\texttt{lake build}, the \texttt{AxiomAudit} step and
per-module \texttt{leanchecker}). The build pins
\texttt{leanprover/lean4:v4.35.0-rc2} because the Mathlib revision tracking
\texttt{master} requires it; no stable release is compatible with that revision
yet. An independent C checker (\leanname{scripts/stab.c}), written for the
internal pre-submission review (see the AI disclosure), is included. It agrees with the Python enumerator on all four certified
obstructions, a further cross-check of the witness edge lists.

The formalization is \emph{logically required} for
Sections~\ref{sec:stab}--\ref{sec:cost}: every numbered theorem, corollary and
lemma there except Theorem~\ref{thm:probe}, and every certified obstruction, is
a theorem in the repository. Five Python scripts are \emph{supporting}, not proof
premises. The exhaustive stability enumerator
(\leanname{scripts/mcs\_stability.py}) reproduces the counterexample witness edge
lists. \leanname{scripts/incomparability\_witness.py} enumerates the MCS and
LexBFS ordering families over all tie-breaks on the witness of
Section~\ref{sec:related} and certifies the $204/204/40$ count. Finally,
\leanname{scripts/mcs\_highn\_crosscheck.py} runs the adopted candidate
implementation (\leanname{scripts/candidates/gen003\_c02.py}) on random graphs up
to $n=200$, checking after every edge flip that its ordering is a legal MCS
ordering and equals a from-scratch recomputation.
\leanname{scripts/prefix\_hypothesis\_count.py} produces the exhaustive counts
after Theorem~\ref{thm:common}, and \leanname{scripts/probe\_check.py} the
random-graph checks of Section~\ref{sec:probe}.

\section{Related work}
\label{sec:related}

The nearest dynamic prior art maintains a property of a \emph{chordal} graph
through an object related to MCS, not the MCS ordering of an arbitrary graph.
Ibarra~\cite{ibarra2008dynamic} maintains a clique-tree representation of a
chordal graph under insertions and deletions. Banerjee, Raman and
Satti~\cite{banerjee2018maintaining} maintain a perfect elimination ordering
under deletions only, with a different window notion and a cell-probe lower
bound for dynamic chordality. Mezzini~\cite{mezzini2012fully} gives
$O(1)$-query/$O(n^2)$-update chordal maintenance. None of these maintains the
MCS ordering itself, and none is machine-checked.

Outside MCS, Morse and Kozsik~\cite{morse2026bfs} maintain a BFS ordering under
edge updates: a framework for fully dynamic BFS spanning trees and their induced
ordering in directed graphs, under both insertions and deletions. They discuss
LexBFS only as a possible refinement and leave a dynamic LexBFS open. It is the
closest sibling to our work, a maintained graph-search ordering of the same
kind, but for BFS rather than MCS, and not machine-checked. Our result is specific to MCS
orderings of undirected graphs, and it is the only one of these that is
machine-checked.

The broader dynamic graph-search literature includes decremental BFS (the
Even--Shiloach tree~\cite{even1981maintaining}), dynamic DFS (Baswana et
al.~\cite{baswana2019dynamic}) and dynamic topological
ordering (Pearce--Kelly~\cite{pearce2007dynamic}; Bender et
al.~\cite{bender2016new}). Among verified dynamic graph algorithms,
the closest is Gu\'eneau et al.~\cite{gueneau2019formal}: incremental cycle
detection in Coq with time credits. Verified SCC algorithms have been formalized
in Isabelle/HOL~\cite{lammich2014gabow}, and Lammich and
Neumann~\cite{lammich2015dfs} give an Isabelle framework for verified DFS. We
found no prior formalization of MCS or of a perfect elimination ordering, and no
machine-checked fully dynamic search-ordering maintenance, in the proof
assistants we surveyed (AFP, Mathlib, Coq libraries, Why3). This is a negative
claim, limited by that survey. MCS-M~\cite{berry2004mcs}, a multi-sweep
extension of MCS for computing minimal triangulations, is static and maintains
no ordering under updates.

The incomparability of MCS and LexBFS orderings is due to
Corneil--Krueger~\cite{corneil2008unified}. On a witness of this
incomparability, a chordal $7$-vertex graph, brute-force enumeration over all
tie-breaks gives $204$ MCS orderings and $204$ LexBFS orderings, and each family
has $40$ orderings not in the other:
$|\mcso\setminus\mathrm{LexBFS}|=|\mathrm{LexBFS}\setminus\mcso|=40$. The witness
is $K_5$ on $\{0,1,2,3,4\}$, plus vertex $5$ joined to $0$ and vertex $6$ joined
to $\{1,2\}$. Brandst\"adt, Dragan and Nicolai~\cite{brandstadt1997lexbfs} show a
common-perfect-elimination-ordering result for LexBFS powers that fails for MCS.
It is adjacent prior art, negative where our stability statement is positive.

\section{Conclusion and open problems}
\label{sec:conclusion}

We have shown how the canonical MCS ordering behaves under single-edge updates:
exact locality, a sublinear first-break probe, an executable refinement, a
probe/pick count, and a stability theorem with certified obstructions for the
set-valued MCS orderings. Everything except the probe bound is machine-checked. Three
directions stand out.

\paragraph{The stability boundary.} Conjecture (Star--Matching): for a one-sided
update $F$, $\mcso(G)\cap\mcso(G\triangle F)\neq\varnothing$ for every host graph
$G$ iff $F$ is a star plus a matching, i.e.\ at most one vertex has $F$-degree
$\ge 2$. The matching side is proved here (Corollary~\ref{cor:matching}); standard
combinatorial reasoning shows a one-sided update $F$ with two vertices of
$F$-degree $\ge 2$ always contains a triangle, a $P_4$ or a $2P_3$ as a subgraph,
of which all three are certified here (Theorem~\ref{thm:obstructions}). However,
subgraph containment does \emph{not} by itself give an obstruction for $F$:
commonness is not monotone in $F$, because adding update edges can restore a
common ordering. The ``only if'' direction therefore needs a \emph{lifting
lemma} (if $F_0\subseteq F$ has a bad host on $n_0$ vertices, then $F$ has a bad
host on some $n\ge n_0$), which is not proved and is open. The star case needs a
non-local inflation argument and is also open.

\paragraph{The adjacent mixed pair.} The disjoint mixed pair is a certified
obstruction. The adjacent mixed pair, where the two flipped edges share a vertex,
never failed in our experiments ($n\le 8$). A lemma proving that it always
preserves a common ordering is the natural next addition.

\paragraph{A maintained deletion probe.} A deletion can first break only at
$\mathrm{pos}(v)$, and with per-step runner-up data $(R_j,r_j)$ the break test
there is $O(1)$ (Section~\ref{sec:probe}). The data $(M_j,w_j)$ alone does not
decide it, and a no-break deletion can change $(R_j,r_j)$ at other steps, so
keeping the runner-up data current in sublinear time is open.

\section*{Acknowledgements and AI disclosure}

This work was developed with AI assistance in accordance with the ELTE IK policy
on the use of AI tools. Three models, each driven through an agent harness over
Lean~4, participated. DeepSeek-V4-Flash-0731 (provider DeepSeek; model
identifier deepseek-ai/DeepSeek-V4-Flash-0731; agentic tool mode, thinking mode
on at default effort) carried the original development phase: the core
definitions, the probe-then-repair updates, and their validity proofs, checked
with lake build over Mathlib. Anthropic Claude Opus 5.5 (provider Anthropic; CLI
coding-agent tool mode, high thinking mode) planned and drafted the
resubmission, wrote the single-source canonical definitions, the locality and
ExecUpdate modules, the stability module, and the unit-cost instrumentation, and
ran the brute-force stability experiments and the revisions answering the
internal reviews. Qwen3.8-27B-FP8 (provider Alibaba; model identifier
Qwen/Qwen3.8-27B-FP8; agentic tool mode, thinking mode on at default effort)
repaired the executable refinement and unit-cost proof bodies, contributed the
certified stability counterexamples, and wired the comparator. In every phase
the model's proof state was machine-checked with lake build and independently
re-checked by the leanchecker external kernel; the transport and bridge lemmas
were written by hand. A self-hosted text-rewriting model (Humanizer 12B, 8-bit
quantized; plain text-completion mode, no reasoning mode) proposed rewordings of
the prose sections and the repository README; every adopted change was applied
by hand and checked against the Lean statements. The internal pre-submission
review rounds, including the C checker scripts/stab.c and the $2P_3$ witness,
were produced by the second author with Anthropic Claude Opus 5.5 (CLI
coding-agent tool mode, high thinking mode) in agent sessions; the probe and prefix-hypothesis check scripts
were written by the first author with the same model, designed from those
review findings. All findings were re-run and checked before being adopted. The human
authors reviewed the resulting definitions, statements, proofs, experiments, and
manuscript text and assume full responsibility for the final work. No AI system
is an author.

\appendix
\section{Core Lean definitions}
\label{app:lean-defs}

The following core definitions, and the Lean statement of
Theorem~\ref{thm:common}, are reproduced verbatim (docstrings and proofs omitted) from the
\leanname{Challenge} module of the artifact (tag \leanname{mcs-arxiv-v10}) so that
readers can audit statement fidelity. The remaining definitions, among them
\leanname{AddsEdges}, \leanname{IsMatching}, \leanname{deleteUpdate} and the
helpers \leanname{firstBreakWindow}, \leanname{earlierPos}, \leanname{laterPos}
and \leanname{greedySuffix} used in \leanname{insertUpdate}, are in the same
module (\texttt{Challenge.lean}).

\begin{lstlisting}[style=lean]
abbrev Vert (n : ℕ) := Fin n

structure UGraph (n : ℕ) where
  adj : Vert n → Vert n → Prop
  symm : ∀ {x y : Vert n}, adj x y → adj y x
  loopless : ∀ (x : Vert n), ¬ adj x x

def neigh (G : UGraph n) (chosen : Finset (Vert n)) (w : Vert n) : ℕ := by
  classical
  exact chosen.card - (chosen.filter (fun x => ¬ G.adj w x)).card

def IsMCSNext (G : UGraph n) (chosen : Finset (Vert n)) (w : Vert n) : Prop :=
  w ∈ (Finset.univ : Finset (Vert n)) \ chosen ∧
  (∀ x : Vert n, x ∈ (Finset.univ : Finset (Vert n)) \ chosen →
    G.neigh chosen x ≤ G.neigh chosen w) ∧
  (∀ x : Vert n, x ∈ (Finset.univ : Finset (Vert n)) \ chosen →
    G.neigh chosen x = G.neigh chosen w → w ≤ x)

def IsMCSOrdering (G : UGraph n) (order : List (Vert n)) : Prop :=
  order.Nodup ∧ order.length = n ∧
  ∀ (i : ℕ) (hi : i < order.length),
    IsMCSNext G (order.take i).toFinset (order.get ⟨i, hi⟩)

def IsMCSNextAny (G : UGraph n) (chosen : Finset (Vert n)) (w : Vert n) : Prop :=
  w ∈ (Finset.univ : Finset (Vert n)) \ chosen ∧
  ∀ x : Vert n, x ∈ (Finset.univ : Finset (Vert n)) \ chosen →
    G.neigh chosen x ≤ G.neigh chosen w

def IsMCSOrderingAny (G : UGraph n) (order : List (Vert n)) : Prop :=
  order.Nodup ∧ order.length = n ∧
  ∀ (i : ℕ) (hi : i < order.length),
    IsMCSNextAny G (order.take i).toFinset (order.get ⟨i, hi⟩)

def IsCommonPrefix (G G' : UGraph n) (P : List (Vert n)) : Prop :=
  P.Nodup ∧ ∀ i (hi : i < P.length),
    IsMCSNextAny G (P.take i).toFinset (P.get ⟨i, hi⟩) ∧
    IsMCSNextAny G' (P.take i).toFinset (P.get ⟨i, hi⟩)

def FlipOf (G G' : UGraph n) (u v : Vert n) : Prop :=
  u ≠ v ∧
  ∀ (x y : Vert n),
    G'.adj x y = G.adj x y ∨ (x = u ∧ y = v) ∨ (x = v ∧ y = u)

def insertUpdate (H : UGraph n) (order : List (Vert n)) (u v : Vert n) : List (Vert n) :=
  let k := Option.getD (firstBreakWindow H order (earlierPos order u v + 1)
    (laterPos order u v)) order.length
  greedySuffix H (order.take k)

theorem exists_common_ordering_prefix : ∀ {n : ℕ} (G G' : UGraph n),
      (∀ (P : List (Vert n)),
          IsCommonPrefix G G' P → P.length < n →
            ∀ (x : Vert n), x ∉ P → G.neigh P.toFinset x ≤ G'.neigh P.toFinset x) →
        (∀ (P : List (Vert n)),
            IsCommonPrefix G G' P → P.length < n →
              ∀ (x : Vert n), x ∉ P → G'.neigh P.toFinset x ≤ G.neigh P.toFinset x + 1) →
          ∃ ord, G.IsMCSOrderingAny ord ∧ G'.IsMCSOrderingAny ord
\end{lstlisting}

The executable layer works over a concrete boolean adjacency matrix
\leanname{Adj}\ $n = \mathsf{List}\ (\mathsf{List}\ \mathsf{Bool})$ with a
transport \leanname{toUGraph}; the instrumented variant
\leanname{execInsertUpdateC} returns the ordering together with the probe/pick
count.

\bibliographystyle{plainnat}
\bibliography{references}

\end{document}